\documentclass[aps,pra,twocolumn,superscriptaddress]{revtex4}
\usepackage{graphics}
\usepackage{subfigure}
\usepackage{amsmath}
\usepackage{physics}
\usepackage{graphicx}
\usepackage{color}
\usepackage{amsfonts}
\usepackage{amssymb}
\usepackage{float}
\usepackage{longtable}
\usepackage{epstopdf}
\usepackage{latexsym}
\usepackage{theorem}
\usepackage{bbm}
\usepackage{hyperref}
\definecolor{darkblue}{rgb}{0,0,0.5}
\hypersetup{
colorlinks=true,
linkcolor=black,
filecolor=blue,
citecolor=darkblue,  
urlcolor=black,
}

\usepackage{bm}
\usepackage{psfrag}
\begin{document}
\title{Idler-free channel position finding}
\author{Jason L. Pereira}
\affiliation{Department of Computer Science, University of York, York YO10 5GH, UK}
\author{Leonardo Banchi}
\affiliation{Department of Physics and Astronomy, University of Florence, via G. Sansone 1, I-50019 Sesto Fiorentino (FI), Italy}
\affiliation{INFN Sezione di Firenze, via G. Sansone 1, I-50019, Sesto Fiorentino (FI), Italy}
\author{Quntao Zhuang}
\affiliation{Department of Electrical and Computer Engineering, University of Arizona, Tucson, Arizona 85721, USA}
\affiliation{James C. Wyant College of Optical Sciences, University of Arizona, Tucson, Arizona 85721, USA}
\author{Stefano Pirandola}
\affiliation{Department of Computer Science, University of York, York YO10 5GH, UK}

\date{\today}

\begin{abstract}
Entanglement is a powerful tool for quantum sensing, and entangled states can greatly boost the discriminative power of protocols for quantum illumination, quantum metrology, or quantum reading. However, entangled state protocols generally require the retention of an idler state, to which the probes are entangled. Storing a quantum state is difficult and so technological limitations can make protocols requiring quantum memories impracticable. One alternative is idler-free protocols that utilise non-classical sources but do not require any idler states to be stored. Here we apply such a protocol to the task of channel position finding. This involves finding a target channel in a sequence of background channels, and has many applications, including quantum sensing, quantum spectroscopy, and quantum reading.
\end{abstract}
\maketitle

\section{Introduction}

Channel position finding (CPF) is a little-investigated but important subcategory of quantum channel discrimination (QCD). In QCD, we know that an unknown channel is drawn from a set of possible channels and our goal is to determine which element of the set it is. In CPF, we have a sequence of channels, all but one of which are identical. The dissimilar channel is the target channel, the remaining channels are background channels, and our goal is to determine the label of the target channel (i.e. find its position in the sequence). This can be expressed as a special case of QCD by considering the entire sequence of channels to be a single multi-channel and the channel sequences given by the different label options to be the elements in the set of possible multi-channels.

Channel position finding is a far less studied task than binary channel discrimination. Discrimination between multiple possible quantum states has been investigated, resulting in, for instance, the development of the pretty good measurement (PGM) \cite{holevo_asymptotically_1978,montanaro_distinguishability_2007}. However, little research has been conducted on the error probability for discriminating between multiple possible quantum channels.

Recently, Zhuang and Pirandola. \cite{zhuang_ultimate_2020} formulated a sequence of lower bounds on the error probability of identifying one channel from a set of possible channels that hold for any set of possible qudit channels and for the most general adaptive protocols, based on channel simulation using port-based teleportation. The bounds were simplified for sets of channels that are jointly teleportation covariant and hence it was shown that the optimal discrimination protocol for such a channel set is non-adaptive. This is a result that was previously only known to hold for binary discrimination \cite{pirandola_ultimate_2017}.

An important case of CPF is locating a (bosonic) thermal loss channel with a different transmissivity or induced noise amongst a sequence of background lossy channels. This is a task with applications in quantum illumination \cite{tan_quantum_2008,karsa_quantum_2020,nair_fundamental_2020}, spectroscopy \cite{mukamel_roadmap_2020,shi_entanglement-assisted_2020}, and quantum reading \cite{pirandola_quantum_2011,pirandola_advances_2018}. In quantum illumination, one may know that a target is present in one of several locations but not know where. A discrimination protocol could involve probing the possible locations with light then collecting and carrying out a measurement on the return states. The different losses and induced noises experienced by the probes, depending on whether they encountered the target or not, could be modelled as different lossy channels. A similar situation could arise in spectroscopy. In this scenario, the different channels could represent the optical absorbance of an unknown substance at different frequencies. Since different substances have different absorption spectra, finding the position of an absorption line could be equivalent to identifying the substance. In quantum reading, the reflectivity of a memory cell takes one of two possible values - encoding one of two possible bit values - and so readout is performed by probing the cell with signal states and discriminating between the possible channels. However, one could also consider a formulation in which bits are instead encoded in the position of a cell with a higher or lower transmissivity than the others \cite{zhuang_entanglement-enhanced_2020}.

Zhuang and Pirandola \cite{zhuang_entanglement-enhanced_2020} upper bounded the performance of classical CPF protocols (meaning non-adaptive protocols for which the signal states have a positive P-representation) and compared them to a specific quantum protocol involving entangled states. This protocol involves sending the signal modes of two-mode squeezed vacuum (TMSV) states through the channels and then measuring them with a proposed new type of receiver called the generalised conditional nulling (CN) receiver. They thereby showed that protocols of this type show a quantum advantage in some regimes.

Despite the advantages of this kind of quantum protocol, storing the idler modes of the probe states is a difficult task. In order to benefit from the use of an idler, it must be stored in a quantum memory, potentially for a long time (if the signal states take a significant amount of time to pass through the channels). Building quantum memories that simultaneously have a long storage time and a high memory efficiency is still a challenging area of research \cite{cho_highly_2016,vernaz-gris_highly-efficient_2018,wang_efficient_2019}. The advantage of idler-free protocols (i.e. quantum protocols that do not involve idlers) is that they can be easier to implement. Hence, the idler-free case is worth investigating, since this will tell us whether we can still achieve a quantum advantage even in the technologically limited case in which we do not have a quantum memory that can store an idler.

GHZ states have been shown to be useful for parameter estimation in quantum metrology, especially for low loss scenarios~\cite{toth_quantum_2014}. This suggests that a continuous variable GHZ state could prove useful for CPF.

In this paper, we start by describing the problem of idler-free channel position finding. We then present the output fidelities (the Bures fidelity between the possible outputs) for the classical and bipartite entangled protocols. We calculate the output fidelity for a protocol using continuous variable GHZ states as probes, which we call the idler-free protocol, and then investigate the behaviour of the various output fidelities, including in the limits of small differences in transmissivity and of large numbers of photons in the probe states. We then introduce what we call the mixed strategy: a protocol that combines aspects of the classical protocol and the idler-free protocol to improve on both (whilst still not requiring an idler). Finally, we apply our results to position-based quantum reading.

\section{Problem specification}

\subsection{General model (quantum pattern recognition)\label{QPRsubsec}}

We are presented with a sequence of $m$ black boxes. Each of these contains a bosonic quantum channel drawn from a known ensemble of possible channels. Let us start by describing the more general problem of quantum pattern recognition, before discussing the specific scenario of CPF that we will be considering here.

Assume we have a label $0\leq i\leq A-1$ and a single-box channel ensemble $\{\mathcal{E}_{i}\}$ spanned by the label ($A$ is the size of the alphabet). In other words, the label of a box specifies which of the $A$ possible channels is in that box. We define a channel pattern over the $m$\ boxes as a sequence $\mathbf{i}:=i_{0},\cdots,i_{m-1}$, with an associated probability $\pi_{\mathbf{i}}$, labelling a multi-channel $\mathcal{E}_{\mathbf{i}}^{m}:=\mathcal{E}_{i_{0}}\otimes\cdots\otimes\mathcal{E}_{i_{m-1}}$. An ensemble of possible patterns is then a set of multi-channels with a corresponding probability distribution, $\{\pi_{\mathbf{i}},\mathcal{E}_{\mathbf{i}}^{m}\}$.

Our input state is an $m$-system probe, $\rho$. This gives rise to an ensemble of possible output states $\{\pi_{\mathbf{i}},\rho_{\mathbf{i}}\}$ where $\rho_{\mathbf{i}}:=\mathcal{E}_{\mathbf{i}}^{m}(\rho)$. Now assume that we probe the pattern $M$ times with $M$ copies of the input state, $\rho^{\otimes M}$, so that the generic output takes the form $\rho_{\mathbf{i}}^{\otimes M}:=\left[  \mathcal{E}_{\mathbf{i}}^{m}(\rho)\right]^{\otimes M}$. We measure this output using a collective positive operator-valued measurement, with measurement operators $\Pi_{\mathbf{j}}$. This defines a specific type of non-adaptive discrimination protocol. We have the associated conditional probabilities
\begin{equation}
p(\mathbf{j}|\mathbf{i}):=\mathrm{Tr}\left(  \Pi_{\mathbf{j}}\rho_{\mathbf{i}
}^{\otimes M}\right)  ,
\end{equation}
and we may define the success probability
\begin{equation}
p_{\text{succ}}(\rho):=\sum_{\mathbf{i}}\pi_{\mathbf{i}}p(\mathbf{i}
|\mathbf{i}),
\end{equation}
or the equivalent error probability $p_{\text{err}}(\rho)=1-p_{\text{succ}}(\rho)$.

We have (a sufficient condition for) quantum advantage when (an upper bound for) the error probability associated to a quantum source is less than (a lower bound for) the error probability associated to a classical source.

For symmetric state discrimination, there are general bounds from Refs.~\cite{bagan_relations_2016,qiu_minimum-error_2010,barnum_reversing_2002,ogawa_strong_1999}. In particular, for any ensemble of mixed states $\{\pi_{\mathbf{i}},\rho_{\mathbf{i}}^{\otimes M}\}$ we may write the following upper bound~\cite{barnum_reversing_2002}
\begin{equation}
p_{\text{err}}\leq\sum_{\mathbf{i}\neq\mathbf{j}}\sqrt{\pi_{\mathbf{i}}
\pi_{\mathbf{j}}}F^{M}(\rho_{\mathbf{i}},\rho_{\mathbf{j}}),\label{eq: Barnum}
\end{equation}
where $F(\sigma_A,\sigma_B)$ is the quantum fidelity, defined by
\begin{equation}
F(\sigma_A,\sigma_B)=\mathrm{Tr}\left[\sqrt{\sqrt{\sigma_A}\sigma_B\sqrt{\sigma_A}}\right],
\end{equation}
and where we have used the multiplicativity of the fidelity over tensor products. Eq.~(\ref{eq: Barnum}) is a bound on the performance of a PGM~\cite{holevo_asymptotically_1978,hausladen_pretty_1994,hausladen_classical_1996}. Then, we may write various lower bounds. In particular, Ref.~\cite{montanaro_lower_2008} derived
\begin{equation}
p_{\text{err}}\geq\frac{1}{2}\sum_{\mathbf{i}\neq\mathbf{j}}\pi_{\mathbf{i}}\pi_{\mathbf{j}}F^{2M}(\rho_{\mathbf{i}},\rho_{\mathbf{j}}).\label{eq: Montanaro}
\end{equation}
These bounds do not depend on the dimension of the channels,$d$. Their proofs hold for any finite $d$ and also for infinite $d$, as long as the states are bona-fide (energy-constrained) quantum states.

If the probability is uniform over a subset of the patterns with size $N$ (and $0$ for all other patterns), then we can set $\pi_{\mathbf{i}}=N^{-1}$ for the patterns with non-zero probability and simplify the bounds using the replacements
\begin{equation}
\sum_{\mathbf{i}\neq\mathbf{j}}\sqrt{\pi_{\mathbf{i}}\pi_{\mathbf{j}}
}\rightarrow N-1,~\sum_{\mathbf{i}\neq\mathbf{j}}\pi_{\mathbf{i}}
\pi_{\mathbf{j}}\rightarrow\frac{N-1}{N}.\label{uniformP}
\end{equation}

\subsection{Idler-free channel position finding}

CPF corresponds to the case in which the single-box channel ensemble is binary, $\{\mathcal{E}_{i}\}=\{\mathcal{E}_{B},\mathcal{E}_{T}\}$, and the patterns are all permutations of $\mathbf{i}:=T,B,\cdots,B$ with equal probability $1/m$. In other words, all of the channels are identical background channels except for a single target channel, and the task is to locate the target channel. For CPF, the pattern $\mathbf{i}$ can equivalently be represented by the position of the target, $i$, so that we can simplify the notation $\{\pi_{\mathbf{i}},\mathcal{E}_{\mathbf{i}}^{m}\}$ into $\{\pi_{i},\mathcal{E}_{i}^{m}\}$, where $\mathcal{E}_{i}^{m}:=\mathcal{E}_{B}^{(0)}\otimes\cdots\otimes\mathcal{E}_{T}^{(i)}\otimes\cdots\otimes\mathcal{E}_{B}^{(m)}$ and $0\leq i\leq m-1$. We denote by $h_i$ the hypothesis that our sequence of channels is the multi-channel $\mathcal{E}_{i}^{m}$. For the specific CPF scenario that we consider here, both the target channel and the background channels are pure-loss channels, albeit with different transmissivities.

Using the previous notation, we have $A=2$ and the probability is uniform over the subset of the patterns that have exactly one occurrence of the label $i=1$. In this case, we can use the replacements in Eq.~(\ref{uniformP}) with $N=m$, so that
\begin{align}
&p_{\text{err}}  \leq(m-1)F^{M}(\rho_{i},\rho_{k}),\label{eq: error prob UB}\\
&p_{\text{err}}  \geq\frac{m-1}{2m}F^{2M}(\rho_{i},\rho_{k}).\label{eq: error prob LB}
\end{align}

The protocols we consider send fully symmetric Gaussian states through the sequence of channels. As per the general model, we use $M$ copies of the input state for the $M$ probes of the sequence of channels, so that the output state has tensor product form (with respect to separate uses of the channel sequence). The input states that we consider are idler-free, meaning we do not retain any modes that are entangled with the signal modes before they are sent through the channels. We allow entanglement between the signal states for each channel, but constrain the total mean number of photons sent through the channel sequence.

\section{Output fidelity calculation}

Due to the tensor product form of the probes that we are using, the key quantity that we must calculate in order to evaluate our bounds on the error probability is the fidelity between possible output states for a single use of the channel sequence (the one-shot output fidelity), $F(\rho_i,\rho_k)$. In fact, if we wish to prove the existence of a quantum advantage using Eqs.~(\ref{eq: error prob UB}) and (\ref{eq: error prob LB}), the sole determinant of whether we will be able to prove that a protocol $\mathcal{A}$ gives a lower error probability than a protocol $\mathcal{B}$ for some number of probes, $M$, is whether the condition
\begin{equation}
F_{\mathcal{A}}(\rho_i,\rho_k)<F^2_{\mathcal{B}}(\rho_i,\rho_k)\label{adv cond}
\end{equation}
holds, where $F_{\mathcal{A}(\mathcal{B})}$ is the output fidelity for protocol $\mathcal{A}$ ($\mathcal{B}$). More specifically, let $p_{err}^{\mathcal{A}(\mathcal{B})}$ be the error probability for protocol $\mathcal{A}$ ($\mathcal{B}$) and let $r_{\mathcal{AB}}$ be the ratio between them. Using the upper bound on $p_{err}^{\mathcal{A}}$ and the lower bound on $p_{err}^{\mathcal{B}}$ (from Eqs.~(\ref{eq: error prob UB}) and (\ref{eq: error prob LB})), we get
\begin{align}
r_{\mathcal{AB}}=\frac{p_{err}^{\mathcal{A}}}{p_{err}^{\mathcal{B}}}\leq 2m \left(\frac{F_{\mathcal{A}}}{F^2_{\mathcal{B}}}\right)^M.
\end{align}
We can immediately see that if the condition in Eq.~(\ref{adv cond}) holds then there exists some value of $M$ beyond which $r_{\mathcal{AB}}<1$ and hence that the upper bound on the error probability of protocol $\mathcal{A}$ is less than the lower bound on the error probability of protocol $\mathcal{B}$.

Note that even if the condition in Eq.~(\ref{adv cond}) does not hold, if the output fidelity for protocol $\mathcal{A}$ is less than the output fidelity for protocol $\mathcal{B}$, this is an indication that protocol $\mathcal{A}$ may outperform protocol $\mathcal{B}$, since the output fidelity is a measure of the distinguishability of the possible outputs.

\subsection{Output fidelity for the classical and bipartite entangled protocols}

There are two protocols that form natural points of comparison for the idler-free protocol (and for the mixed strategy): the optimal classical protocol (i.e. the optimal non-adaptive protocol that uses probes with a positive P-representation) and a protocol that uses a tensor product of $m$ individual TMSV states as the input and sends one mode of a TMSV state through each channel, retaining the other mode as an idler (which we will call the bipartite entangled protocol). If our idler-free protocol is no better than the best classical protocol, this suggests that there is little benefit to using a quantum source. If the bipartite entangled protocol has a performance that is similar to that of the idler-free protocol, this is an indication that we lose little in terms of performance by limiting ourselves to idler-free protocols.

Let us start by giving the output fidelity and a lower bound on the error probability for the optimal classical protocol (which we will henceforth simply refer to as the classical protocol), as calculated in Ref.~\cite{zhuang_entanglement-enhanced_2020}. Lemma~4 of Ref.~\cite{zhuang_entanglement-enhanced_2020} assumes a global energy constraint of $mMN_{S}$ mean photons, with $M$ modes
irradiated over each of the $m$ boxes. $N_S$ is the average number of photons sent through a channel per channel use. The minimum output fidelity for a classical, non-adaptive protocol is
\begin{equation}
F^{\mathrm{class}}=\exp\left[-MN_{S}(\sqrt{\eta_{B}}-\sqrt{\eta_{T}})^{2}\right],\label{eq: fid_class}
\end{equation}
where $\eta_{B(T)}$ is the transmissivity of the background (target) channel, which has no dependence on the number of channels. An optimal probe saturating this bound is the tensor-product state $\otimes_{k=1}^{m}\ket{\sqrt{MN_S}}_{k}$. The error probability for such a protocol is then lower bounded by
\begin{equation}
p_{\text{err}}^{\text{class}}\geq\frac{m-1}{2m}\exp\left[-2MN_{S}
(\sqrt{\eta_{B}}-\sqrt{\eta_{T}})^{2}\right],\label{LB_class}
\end{equation}
which holds with no restriction on the number of modes irradiated per box (so long as the energy constraint holds).

We can also give an achievable upper bound on the error probability, in order to give a point of comparison for other protocols. For pure output states (which we obtain, since all of the channels in our ensemble are pure-loss channels), we can use the PGM to achieve the error bound~\cite{cariolaro_theory_2010,zhuang_entanglement-enhanced_2020}
\begin{equation}
p_{\text{err}}^{\text{class}}\leq \frac{m-1}{m^2}(\sqrt{1+(m-1)F^{\mathrm{class}}}-\sqrt{1-F^{\mathrm{class}}})^2.
\end{equation}

Let us consider the bipartite entangled protocol. In this case, the input state for each channel is the TMSV state $\Phi_{\mu}$, with the first mode retained as an idler and the second mode probing the box. The covariance matrix (CM) of $\Phi_{\mu}$ (for vacuum noise $1$) is
\begin{equation}
V=\begin{pmatrix}
\mu \mathbb{I} & \mu^{\prime}\mathbb{Z}\\
\mu^{\prime}\mathbb{Z} & \mu \mathbb{I}
\end{pmatrix},~
\begin{array}
[c]{l}
\mu=2N_{S}+1,\\
\mu^{\prime}=\sqrt{\mu^{2}-1},
\end{array}
\end{equation}
where $N_S$ is the mean number of photons sent through each channel (and is the same for every box) and $\mathbb{Z}$ is the Pauli Z-matrix.

Note that in this (and the classical) case, the possible output states can all be written as tensor products of the output states of each channel. Any two possible outputs (for a single use of the channel sequence), $\rho_i$ and $\rho_j$ (where the subscript, as usual, labels the target position), will differ in only two subsystems: the subsystems corresponding to the individual probes used for the $i$-th and $j$-th channels. Due to the multiplicativity of the fidelity with respect to tensor products, the output fidelity can be calculated taking into account only these two subsystems (but including both the signal states and their respective idlers).

Tracing out the other subsystems, we can write the reduced output states, $\rho_{i}'$ and $\rho_{j}'$, as
\begin{align}
\rho_{i}'&=(\openone\otimes \mathcal E_T)(\openone\otimes \mathcal E_B)[\Phi_\mu^{\otimes 2}] = \chi_{\mathcal E_T}^\mu \otimes \chi_{\mathcal E_B}^\mu,\\
\rho_{j}'&=(\openone\otimes \mathcal E_B)(\openone\otimes \mathcal E_B)[\Phi_\mu^{\otimes 2}]
= \chi_{\mathcal E_B}^\mu \otimes \chi_{\mathcal E_T}^\mu,
\end{align}
where $\chi_{\mathcal E}^\mu$ is the finite energy approximation of the Choi matrix of channel $\mathcal E$.
We can therefore calculate the output fidelity using
\begin{equation}
F(\rho_{i},\rho_{j})=F(\chi_{\mathcal E_T}^\mu, \chi_{\mathcal E_B}^\mu)^2,
\end{equation}
where the covariance matrices of $\chi_{\mathcal E_B}^\mu$ and $\chi_{\mathcal E_T}^\mu$ are
\begin{align}
&V_{\chi_{\mathcal E_B}^\mu} =\begin{pmatrix}
\mu \mathbb{I} & \sqrt{\eta_{B}}\mu^{\prime}\mathbb{Z}\\
\sqrt{\eta_{B}}\mu^{\prime}\mathbb{Z} & (\eta_{B}\mu+(1-\eta_{B}))\mathbb{I}
\end{pmatrix},\\
&V_{\chi_{\mathcal E_T}^\mu} =\begin{pmatrix}
\mu \mathbb{I} & \sqrt{\eta_{T}}\mu^{\prime}\mathbb{Z}\\
\sqrt{\eta_{T}}\mu^{\prime}\mathbb{Z} & (\eta_{T}\mu+(1-\eta_{T}))\mathbb{I}
\end{pmatrix}.
\end{align}
The resulting formula for the output fidelity is
\begin{equation}
F^{\mathrm{bipartite}}=(1+N_S(1-\sqrt{(1-\eta_B)(1-\eta_T)}-\sqrt{\eta_B \eta_T}))^{-2},\label{eq: bipartite ent fid}
\end{equation}
which again has no dependence on $m$.

\subsection{Output fidelity for the idler-free protocol}

We now move on to the idler-free protocol. In this case, our probe, $\Phi_{\mu}^{m}$, is an $m$-partite, fully symmetric Gaussian state. The mean photon number per channel is $N_{S}$. The input state, $\rho_{\mathrm{in}}$, has the CM
\begin{equation}
V_{\mathrm{in}}=\begin{pmatrix}
\mu \mathbb{I} & \Gamma & \dots & \Gamma\\
\Gamma & \mu I & \ddots & \Gamma\\
\vdots & \ddots & \ddots & \vdots\\
\Gamma & \Gamma & \dots & \mu \mathbb{I}
\end{pmatrix},
~\Gamma:=\mathrm{diag}(c_{1},c_{2}),\label{eq: idler-free input}
\end{equation}
which has a simple symplectic spectrum, namely the two
eigenvalues~\cite{serafini_unitarily_2005}
\begin{align}
\nu_{-} &  =\sqrt{(\mu-c_{1})(\mu-c_{2})}\\
\nu_{+} &  =\sqrt{[\mu+(m-1)c_{1}][\mu+(m-1)c_{2}]}
\end{align}
where $\nu_{+}$ is $m-1$ times degenerate. We will henceforth assume $c_{1}=-c_{2}:=c$. From the
bona fide condition $\nu_{\pm}\geq1$ (and $V>0$) we can derive
\begin{equation}
|c|\leq\frac{\sqrt{\mu^{2}-1}}{m-1}.
\end{equation}
Let us assume maximal correlations:
\begin{equation}
c_{\text{max}}=(m-1)^{-1}\sqrt{\mu^{2}-1}.\label{cmax}
\end{equation}
Note that, for $c=c_{\text{max}}$, $\nu_{+}=1$, but $\nu_{-}>1$ unless $m=2$. Consequently, $\Phi_{\mu}^{m}$ is not pure.

The first mode probes the first box, the second mode probes the second box, etc. Under hypothesis $h_{i}$, i.e. for the multiple pure-loss channel
\begin{equation}
\mathcal{E}_{i}^{m}:=\mathcal{E}^{(0)}(\eta_{B})\otimes\cdots\otimes
\mathcal{E}^{(i)}(\eta_{T})\otimes\cdots\otimes\mathcal{E}^{(m-1)}(\eta_{B}),
\end{equation}
we have the output CM
\begin{equation}
V_{i}=\begin{pmatrix}
\Delta_{B} & \cdots & \Gamma_{B} & \Gamma_{T} & \Gamma_{B} & \cdots &\Gamma_{B}\\
\vdots & \ddots & \vdots & \vdots & \vdots &  & \vdots\\
\Gamma_{B} &  & \Delta_{B} & \Gamma_{T} & \Gamma_{B} & \cdots & \Gamma_{B}\\
\Gamma_{T} & \cdots & \Gamma_{T} & \Delta_{T} & \Gamma_{T} & \cdots &\Gamma_{T}\\
\Gamma_{B} & \cdots & \Gamma_{B} & \Gamma_{T} & \Delta_{B} &  & \Gamma_{B}\\
\vdots &  &  & \vdots &  & \ddots & \vdots\\
\Gamma_{B} & \cdots & \Gamma_{B} & \Gamma_{T} & \Gamma_{B} & \cdots &\Delta_{B}
\end{pmatrix},\label{CMoutput}
\end{equation}
where we have defined
\begin{align}
\Delta_{B} &  :=(\eta_{B}\mu+(1-\eta_{B}))\mathbb{I},~\Delta_{T}:=(\eta_{T}\mu+(1-\eta_{T}))\mathbb{I},\label{eq: output CM delta}\\
\Gamma_{B} & =\eta_{B}c_{\text{max}}\mathbb{Z},~\Gamma_{T}  =\sqrt{\eta_{B}\eta_{T}}c_{\text{max}}\mathbb{Z}.\label{eq: output CM gamma}
\end{align}

For the binary case ($m=2$), our two possible output CMs are
\begin{align}
V_{0}=\begin{pmatrix}
\Delta_{T} & \Gamma_{T}\\
\Gamma_{T} & \Delta_{B}
\end{pmatrix},~
V_{1}=\begin{pmatrix}
\Delta_{B} & \Gamma_{T}\\
\Gamma_{T} & \Delta_{T}
\end{pmatrix}.
\end{align}
Then, applying the formula for the fidelity of a Gaussian state from Ref.~\cite{banchi_quantum_2015}, we get
\begin{equation}
\begin{split}
F^{\mathrm{idler-free},\mathrm{binary}}=&(1+N_S(\eta_B+\eta_T-2\eta_B\eta_T\\
&-2\sqrt{\eta_B\eta_T(1-\eta_B)(1-\eta_T)}))^{-1}.\label{eq: idler free binary fid}
\end{split}
\end{equation}
The output fidelity for the $m=3$ case is calculable in a similar way. 

Let us now address the more complicated case of $m>3$. Whilst the formulae in Ref.~\cite{banchi_quantum_2015} apply for any number of modes, the calculations involved become much more difficult for large $m$. Therefore, it is helpful to reduce the calculation of the fidelity between two $m$-mode states to a far more tractable calculation of the fidelity between two $3$-mode states.

\subsubsection{Reduction of the output fidelity calculation}
$F(V_{i},V_{j})=F(V_{1},V_{2})$ for all $i\neq j$, so it suffices to calculate $F(V_{1},V_{2})$. Let $\{ a _i\}$ be the set of annihilation operators for all of the modes. We can transform $\{ a _i\}$ via the unitary
\begin{align}
U=\mathbb{I}_{1,2}\otimes U^{\prime}
\end{align}
where $\mathbb{I}_{1,2}$ is the identity on modes 1 and 2 and where $U'$ has elements
\begin{align}
&U^{\prime}_{jk}=\frac{e^{i j k \phi}}{\sqrt{m-2}},~\phi=\frac{2\pi}{m-2}.
\end{align}
We can verify that $U^{\prime}$ is a valid unitary by writing
\begin{align}
(U^{\prime}U^{\prime\dagger})_{jk}&=\frac{1}{m-2}\sum_{l=1}^{m-2}e^{i(k-j)l\phi}\\
&=\delta_{jk}.
\end{align}
$U$ transforms $\{ a _i\}$ into $\{ a ^{\prime}_i\}$, where
\begin{align}
& a ^{\prime}_1= a _1,~ a ^{\prime}_2= a _2,\\
& a ^{\prime}_{3+j}=\frac{1}{\sqrt{m-2}}\sum_{k=0}^{m-3}e^{i k\phi} a _{3+k}.
\end{align}
This means that the quadrature operators of the modes, $\{q_i\}$ and $\{ p _i\}$, are transformed into $\{q^{\prime}_i\}$ and $\{ p ^{\prime}_i\}$, where
\begin{align}
& q ^{\prime}_1= q _1,~ q ^{\prime}_2= q _2,\\
& p ^{\prime}_1= p _1,~ p ^{\prime}_2= p _2,\\
& q ^{\prime}_{3+j}=\frac{1}{\sqrt{m-2}}\sum_{k=0}^{m-3}\left[\cos{(j k\phi)} q _{3+k}-\sin{(j k\phi)} p _{3+k}\right],\\
& p ^{\prime}_{3+j}=\frac{1}{\sqrt{m-2}}\sum_{k=0}^{m-3}\left[\sin{(j k\phi)} q _{3+k}+\cos{(j k\phi)} p _{3+k}\right].\label{eq: transformed p}
\end{align}
These are calculated using the relations $ q = a + a ^{\dagger}$ and $ p =i( a ^{\dagger}- a )$.

This transformation puts both $V_1$ and $V_2$ in block diagonal form, such that the resulting CM has a 6 by 6 block and a $2m-6$ by $2m-6$ block, the latter of which is the same in both cases. We can verify this by calculating the components of the transformed CMs, $V_{1}^{\prime}$ and $V_{2}^{\prime}$. In order to demonstrate how this is done, let us explicitly calculate the value of $\left\langle  p ^{\prime}_{1} p ^{\prime}_{3+j}  \right\rangle$ for $V_{1}^{\prime}$. Using the expression in Eq.~(\ref{eq: transformed p}), we get
\begin{align}
\begin{split}
\left\langle  p ^{\prime}_{1} p ^{\prime}_{3+j} \right\rangle = \frac{1}{\sqrt{m-2}}&\sum_{k=0}^{m-3}\left[\sin{(j k\phi)}\left\langle p _{1} q _{3+k}\right\rangle\right.\\
&\left.+\cos{(j k\phi)}\left\langle p _{1} p _{3+k}\right\rangle\right].
\end{split}
\end{align}
The covariances $\left\langle p _{1} q _{3+k}\right\rangle$ and $\left\langle p _{1} p _{3+k}\right\rangle$ are components of the original covariance matrix, $V_1$, and are given in Eqs.~(\ref{CMoutput}) to (\ref{eq: output CM gamma}). First, note that $\left\langle  p _{i} q _{j} \right\rangle=0$ for all $i$ and $j$. We now define
\begin{align}
&d_B=\eta_B\mu+(1-\eta_B),~d_T=\eta_T\mu+(1-\eta_T),\label{eq: def delta}\\
&\gamma_B=\eta_B c_{\text{max}},~\gamma_T=\sqrt{\eta_B \eta_T}c_{\text{max}}.\label{eq: def gamma}
\end{align}
Hence, we obtain
\begin{align}
\begin{split}
\left\langle  p ^{\prime}_{1} p ^{\prime}_{3+j} \right\rangle &= -\frac{\gamma_T}{\sqrt{m-2}}\sum_{k=0}^{m-3}\cos{(j k\phi)}\\
&=-\sqrt{m-2}\gamma_T\delta_{0,j},
\end{split}
\end{align}
where $\delta$ is the Kronecker delta symbol, and where we have used the result
\begin{align}
\sum_{k=1}^{n}\cos{(j k\frac{2\pi}{n})}=n\delta_{0,j},
\end{align}
for integer $n$. Note that this is 0 for $j>0$, i.e. for all modes with labels greater than 3. $\left\langle  q ^{\prime}_{1} q ^{\prime}_{3+j} \right\rangle$ is simply $-\left\langle  p ^{\prime}_{1} p ^{\prime}_{3+j} \right\rangle$, and $\left\langle  p ^{\prime}_{2} p ^{\prime}_{3+j} \right\rangle$ can be obtained simply by substituting $\eta_{B}c_{\text{max}}$ for $\sqrt{\eta_{B}\eta_{T}}c_{\text{max}}$, giving
\begin{align}
\left\langle p ^{\prime}_{2} p ^{\prime}_{3+j} \right\rangle = -\sqrt{m-2}\gamma_B\delta_{0,j}.
\end{align}
Note that for $V_2$, we simply swap $\left\langle  p ( q )^{\prime}_{1} p ( q )^{\prime}_{3+j} \right\rangle$ and $\left\langle  p ( q )^{\prime}_{2} p ( q )^{\prime}_{3+j} \right\rangle$.

We have now shown that no correlations exist between modes 1 and 2 and modes 4 to $m$. In order to show that the transformation puts the CM in block diagonal form, we must also show that no correlations exist between mode 3 and modes 4 to $m$. To do this, we must calculate $\left\langle  p ( q )^{\prime}_{3} p ( q )^{\prime}_{3+j} \right\rangle$. Again using Eq.~(\ref{eq: transformed p}), we obtain
\begin{align}
\begin{split}
\left\langle  p ^{\prime}_{3} p ^{\prime}_{3+j} \right\rangle = \frac{1}{m-2}&\sum_{k,l=0}^{m-3}\left[\sin{(j k\phi)}\left\langle p _{3+l} q _{3+k}\right\rangle\right.\\
&\left.+\cos{(j k\phi)}\left\langle p _{3+l} p _{3+k}\right\rangle\right].
\end{split}
\end{align}
Substituting in Eqs.~(\ref{eq: def delta}) and (\ref{eq: def gamma}), we derive
\begin{align}
\begin{split}
\left\langle  p ^{\prime}_{3} p ^{\prime}_{3+j} \right\rangle = &-\gamma_B\sum_{k=0}^{m-3}\cos{(j k\phi)}+\frac{d_B+\gamma_B}{m-2}\sum_{k=0}^{m-3}\cos{(j k\phi)},
\end{split}
\end{align}
where we have split the expression into contributions from the on and off-diagonal components of the original CMs. Simplifying, we get
\begin{align}
\left\langle  p ^{\prime}_{3} p ^{\prime}_{3+j} \right\rangle = (d_B-(m-3)\gamma_B)\delta_{0,j},
\end{align}
thus there are no correlations between mode 3 and modes 4 to $m$. We have therefore carried out a unitary transform on $V_1$ and $V_2$ such that they are in block diagonal form, with a 6 by 6 block and a $2m-6$ by $2m-6$ block. Since the $2m-6$ by $2m-6$ block is the same for both $V_1$ and $V_2$, we can ignore this block (trace over the remaining $m-3$ modes) when calculating the fidelity of the two CMs. This reduces the problem to the analytically solvable case of finding the fidelity of a pair of three-mode Gaussian states.

Let $V_1^{\prime}$ be the CM of $\rho_1$ after the unitary $U$ has been enacted on it, transforming it into block diagonal form. Then, let $V_1^{\prime\prime}$ be the CM after the last $m-3$ modes have been discarded. $V_1^{\prime\prime}$ takes the form
\begin{align}
V_1^{\prime\prime}=\begin{pmatrix}
\Delta_T &\Gamma_T &\sqrt{m-2}\Gamma_T\\
\Gamma_T &\Delta_B &\sqrt{m-2}\Gamma_B\\
\sqrt{m-2}\Gamma_T &\sqrt{m-2}\Gamma_B &\Delta_B+(m-3)\Gamma_B
\end{pmatrix}.
\end{align}
To obtain $V_2^{\prime\prime}$, we simply swap modes 1 and 2.

We can also calculate the structure of the traced over modes, although this does not affect the fidelity calculation, since it is the same for both $V^{\prime}_1$ and $V^{\prime}_2$. Let us calculate $\left\langle  p ^{\prime}_{3+j} p ^{\prime}_{3+k} \right\rangle$ for $j,k>0$. Considering only the non-zero components, we get
\begin{align}
\begin{split}
\left\langle  p ^{\prime}_{3+j} p ^{\prime}_{3+k} \right\rangle = \frac{1}{m-2}&\sum_{x,y=0}^{m-3}\left[\sin{(j x\phi)}\sin{(k y\phi)}\left\langle q _{3+x} q _{3+y}\right\rangle\right.\\
&\left.+\cos{(j x\phi)}\cos{(k y\phi)}\left\langle p _{3+x} p _{3+y}\right\rangle\right].
\end{split}
\end{align}
We now split this into three terms, by writing
\begin{align}
&\left\langle  p ^{\prime}_{3+j} p ^{\prime}_{3+k} \right\rangle = \frac{t_1+t_2+t_3}{m-2},\\
&t_1=\gamma_B \sum_{x,y=0}^{m-3} \left[\sin{(j x\phi)}\sin{(k y\phi)}-\cos{(j x\phi)}\cos{(k y\phi)}\right],\\
&t_2=d_B \sum_{x=0}^{m-3} \left[\sin{(j x\phi)}\sin{(k x\phi)}+\cos{(j x\phi)}\cos{(k x\phi)}\right],\\
&t_3=-\gamma_B \sum_{x=0}^{m-3} \left[\sin{(j x\phi)}\sin{(k x\phi)}-\cos{(j x\phi)}\cos{(k y\phi)}\right].
\end{align}
We can then write
\begin{align}
&t_1=-\gamma_B \sum_{x,y=0}^{m-3} \cos{((jx+ky)\phi)},\\
&t_2=d_B \sum_{x=0}^{m-3} \cos{((j-k) x\phi)},\\
&t_3=\gamma_B \sum_{x=0}^{m-3} \cos{((j+k) x\phi)},
\end{align}
where we have used
\begin{align}
\cos{(a+b)=\cos{(a)}\cos{(b)}-\sin{(a)}\sin{(b)}}.
\end{align}
Since $j,k>0$, $t_1=0$. $t_2$ is non-zero iff $j=k$ and $t_3$ is non-zero iff $j+k=m-2$. We therefore derive
\begin{align}
\left\langle  p ^{\prime}_{3+j} p ^{\prime}_{3+k} \right\rangle = d_B \delta_{j,k}+\gamma_B \delta_{j+k,m-2}.
\end{align}
Via a similar derivation, we find
\begin{align}
\left\langle  q ^{\prime}_{3+j} q ^{\prime}_{3+k} \right\rangle = d_B \delta_{j,k}-\gamma_B \delta_{j+k,m-2}.
\end{align}
We now have all of the components of the CM of the traced over modes.

The fidelity $F(V_1,V_2)=F(V_1^{\prime\prime},V_2^{\prime\prime})$ can now be easily found using the formula from \cite{banchi_quantum_2015}.

\section{Behaviour of the output fidelity}

If the fidelity between the possible output states (the output fidelity) of the idler-free protocol is lower than that of the classical protocol over some parameter range, it is an indication that there is a benefit to using the input state described by Eq.~(\ref{eq: idler-free input}), rather than using the classical protocol. If the output fidelity for the idler-free protocol is close to that of the entangled state protocol with idlers, this indicates that the cost to performance of using an idler-free protocol is small.

Note that this is only an indication, as the fidelity is a measure of the distinguishability of states, but this does not necessarily mean that the error probability in discriminating between states is completely determined by the fidelity. As mentioned previously, in order to prove an advantage of one protocol over another, we have to compare upper and lower bounds on the error probability. Nonetheless, since our bounds on the error probability in Eqs.~(\ref{eq: error prob UB}) and (\ref{eq: error prob LB}) depend on the output fidelity, it makes sense to compare the behaviours of the output fidelity functions for the classical, bipartite entangled, and idler-free cases.

\subsection{Behaviour with respect to the number of channels in the sequence}

\begin{figure}[tb]
\centering
\includegraphics[width=1\linewidth]{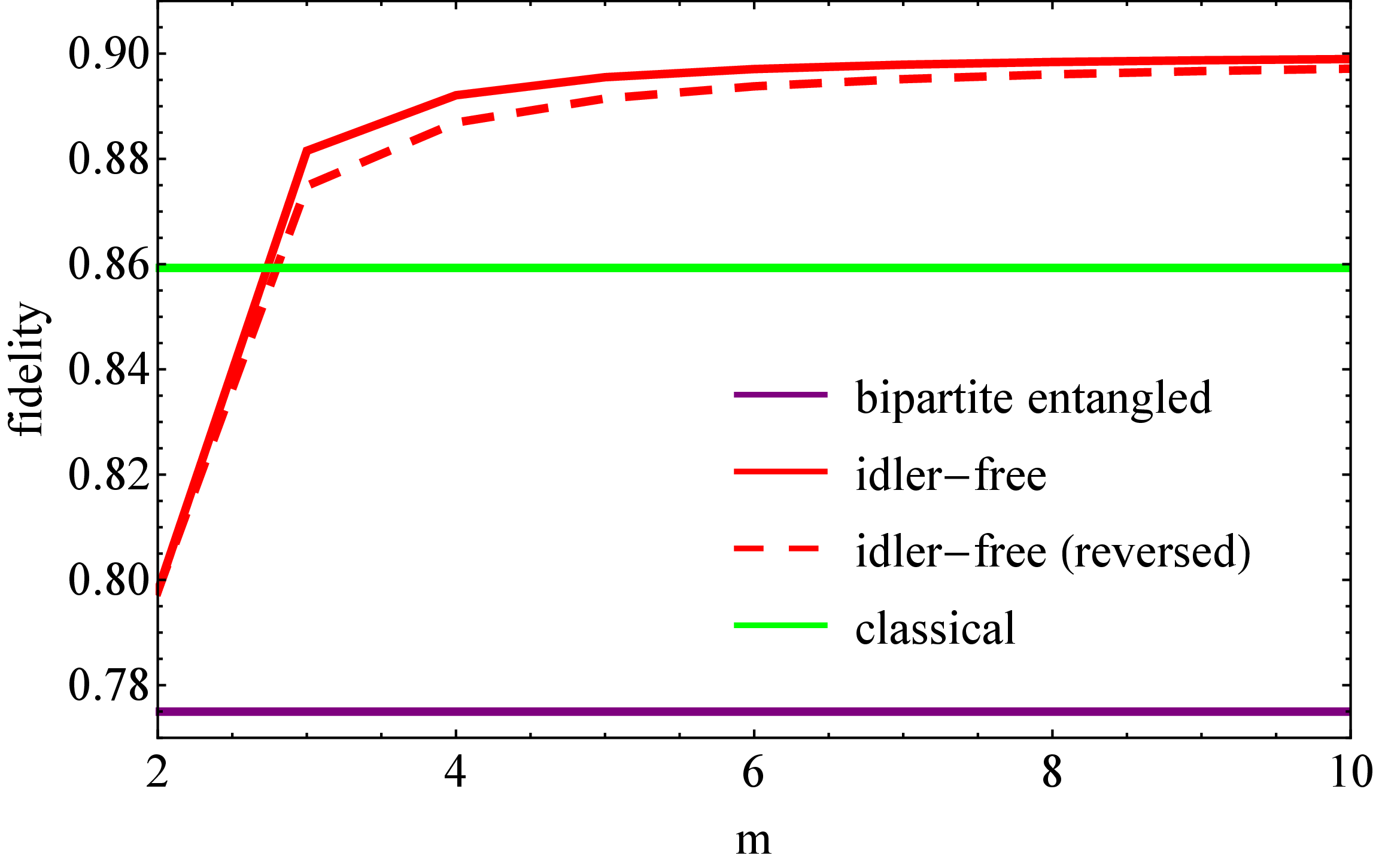}\caption{The output fidelity of the classical, bipartite entangled, and idler-free protocols as a function of the total number of channels in the sequence, $m$. We set the background transmissivity, $\eta_B=0.2$, the transmissivity of the target channel, $\eta_T=0.7$, and the average number of photons per channel use, $N_S=1$. Only the idler-free protocol is affected by changing $m$. We see that the output transmissivity increases as $m$ increases, but levels off for large $m$. As $m$ increases, the effect on the output fidelity of swapping $\eta_B$ and $\eta_T$ decreases.}
\label{fig:idler free vs m}
\end{figure}

Fig.~\ref{fig:idler free vs m} plots the output fidelities for the various protocols against the number of channels in the sequence. In this plot, $\eta_B=0.2$, $\eta_T=0.7$, and $N_S=1$. As previously mentioned, the output fidelities of the classical and the bipartite entangled protocols do not depend on $m$. Fig.~\ref{fig:idler free vs m} shows that the output fidelity for the idler-free protocol increases as $m$ increases, but levels off for large $m$.

Figs.~\ref{fig:idler free vs m}, ~\ref{fig:idler free vs eta}, and \ref{fig:idler free vs eta, low eta} also have a curve labelled ``idler-free (reversed)". This gives the fidelity for the idler-free protocol when the values of $\eta_B$ and $\eta_T$ are swapped. It is immediate from the tensor product structure of the outputs that neither the fidelity of the classical protocol nor that of the bipartite entangled protocol are affected by swapping $\eta_B$ and $\eta_T$, however this is not the case for the idler-free protocol (for $m>2$). In fact, Figs.~\ref{fig:idler free vs eta} and \ref{fig:idler free vs eta, low eta} show that there can be a significant difference between the two fidelities.

Since the output fidelity for the idler-free protocol increases with $m$, the $m=2$ scenario is an important case to study when comparing the protocols. This output fidelity for this scenario is given in Eq.~(\ref{eq: idler free binary fid}).

\subsection{Behaviour with respect to the transmissivities}

\begin{figure}[tb]
\centering
\includegraphics[width=1\linewidth]{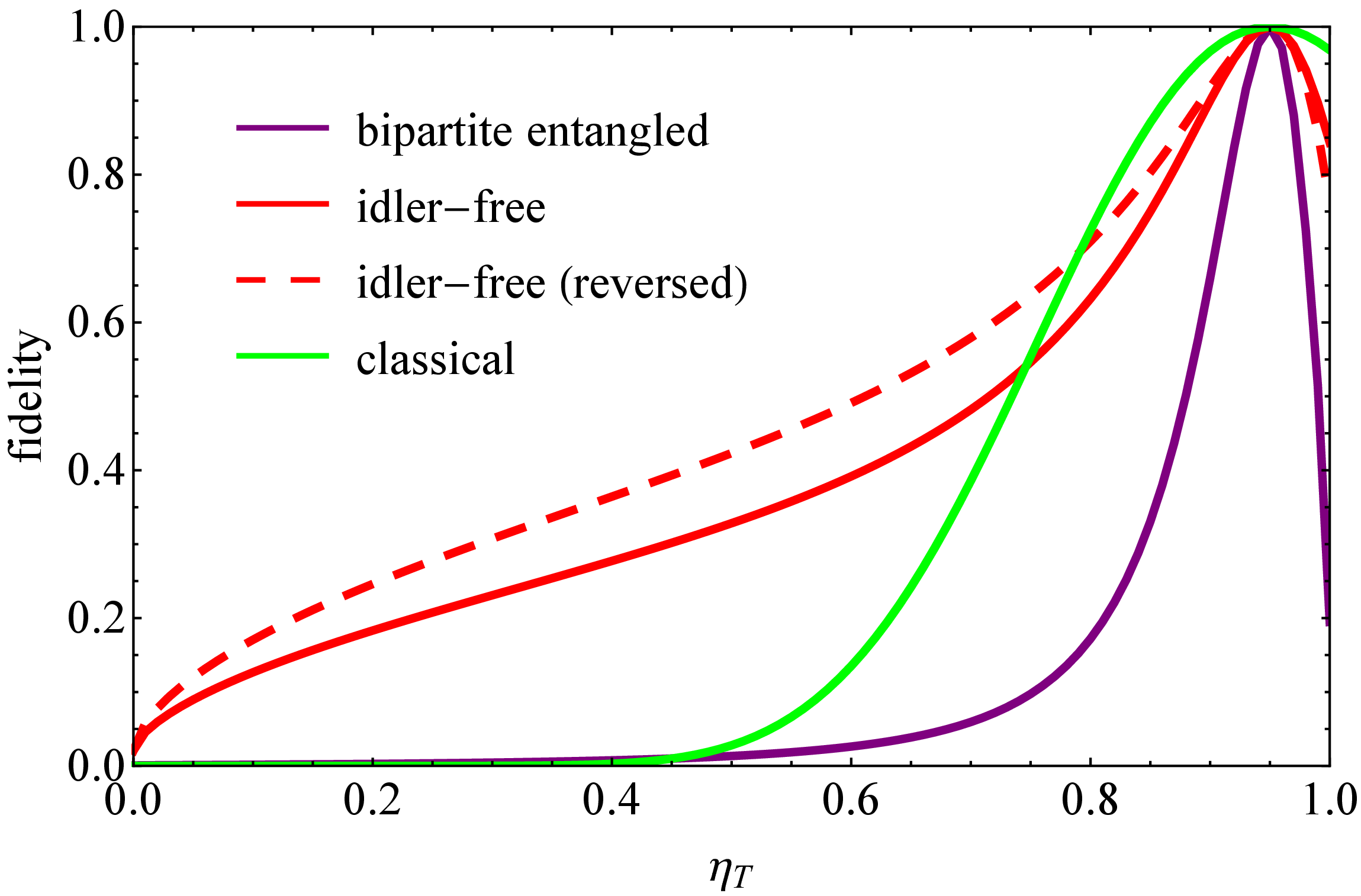}\caption{The output fidelity of the classical, bipartite entangled, and idler-free protocols as a function of the transmissivity of the target channel, $\eta_T$.  We set the transmissivity of the background channels, $\eta_B=0.95$, and impose an energy constraint so that the average number of photons per channel use is no more than $50$. We also set $m=3$, so that there are two identical background channels and one target channel. The output fidelity for the idler-free protocol with $\eta_B$ and $\eta_T$ swapped is also shown. Unlike for the classical and bipartite entangled protocols, this swap affects the output fidelity for the idler-free protocol. The output fidelities are highest when $\eta_T$ is close to $\eta_B$ and decrease as the difference between the two transmissivities increases. The idler-free protocol gives a lower output fidelity than the classical protocol for $\eta_T\gtrapprox0.75$.}
\label{fig:idler free vs eta}
\end{figure}

In Fig.~\ref{fig:idler free vs eta}, we plot the various output fidelities against the transmissivity of the target channel, $\eta_T$. We fix the background transmissivity, $\eta_B=0.95$, the number of channels in the sequence, $m=3$, and the average number of photons sent through each channel per channel use, $N_S=50$. We see that there is a region ($\eta_T\gtrapprox0.75$) for which the idler-free protocol has a lower fidelity than the classical protocol. This indicates that the idler-free protocol could have a use as an intermediate between the easily implemented classical protocol, based on the sending of coherent states, and the bipartite entangled protocol, which gives a lower output fidelity in this range but could be harder to implement, due to the need for a quantum memory to preserve the idlers. The idler-free protocol could be easier to implement, despite the fact it still requires the generation of a non-classical state, because it does not require a quantum memory.

We can investigate how sensitive the output fidelity functions are for small differences in the transmissivities of the target and background channels by setting $\eta_T=\eta$ and $\eta_B=\eta+\epsilon$ and then expanding the functions in terms of $\epsilon$. For the idler-free protocol in the binary case ($m=2$), we get
\begin{equation}
F^{\mathrm{idler-free},\mathrm{binary}}\approx 1-\frac{N_S}{4\eta(1-\eta)}\epsilon^2+\mathcal{O}(\epsilon^3),
\end{equation}
where we have expanded up to the second power in $\epsilon$. For the classical protocol, we get
\begin{equation}
F^{\mathrm{class}}\approx 1-\frac{N_S}{4\eta}\epsilon^2+\mathcal{O}(\epsilon^3),
\end{equation}
and for the bipartite entangled protocol, we get
\begin{equation}
F^{\mathrm{bipartite}}\approx 1-\frac{N_S}{4\eta(1-\eta)}\epsilon^2+\mathcal{O}(\epsilon^3).
\end{equation}
Note that the expansion for the idler-free protocol, up to the $\epsilon^2$ term, is the same as for the bipartite entangled protocol. In fact, the expansions only differ at the $\epsilon^4$ term and above, suggesting that the idler-free protocol may be a good substitute for the bipartite entangled protocol in scenarios in which the target and background channels have similar transmissivities.

Note that the $\epsilon^2$ term for the classical protocol has an extra multiplicative factor of $(1-\eta)$ compared to the $\epsilon^2$ term for the (binary) idler-free and bipartite entangled protocols, and so these protocols achieve lower output fidelities than the classical protocol for sufficiently small $\epsilon$. This also suggests that the relative sensitivities of the idler-free and bipartite entangled protocols (compared to the classical protocol), for small $\epsilon$, are greater for high $\eta$ (i.e. when both the target and the background channels have high transmissivities).

We expect the idler-free protocol to perform worse than the classical protocol for low $\eta_B$. The reason for this can be seen by considering the extreme scenario in which $\eta_B=0$ (and $\eta_T>0$). Then, the background modes are lost. We are left with the target mode, which is in a thermal state that has passed through the lossy channel $\mathcal{E}_T$. In other words, the output could be equivalently obtained by sending a tensor product of thermal states through the channel sequence. Consequently, in this scenario, the idler-free protocol cannot outperform the classical protocol.

\begin{figure}[tb]
\centering
\includegraphics[width=1\linewidth]{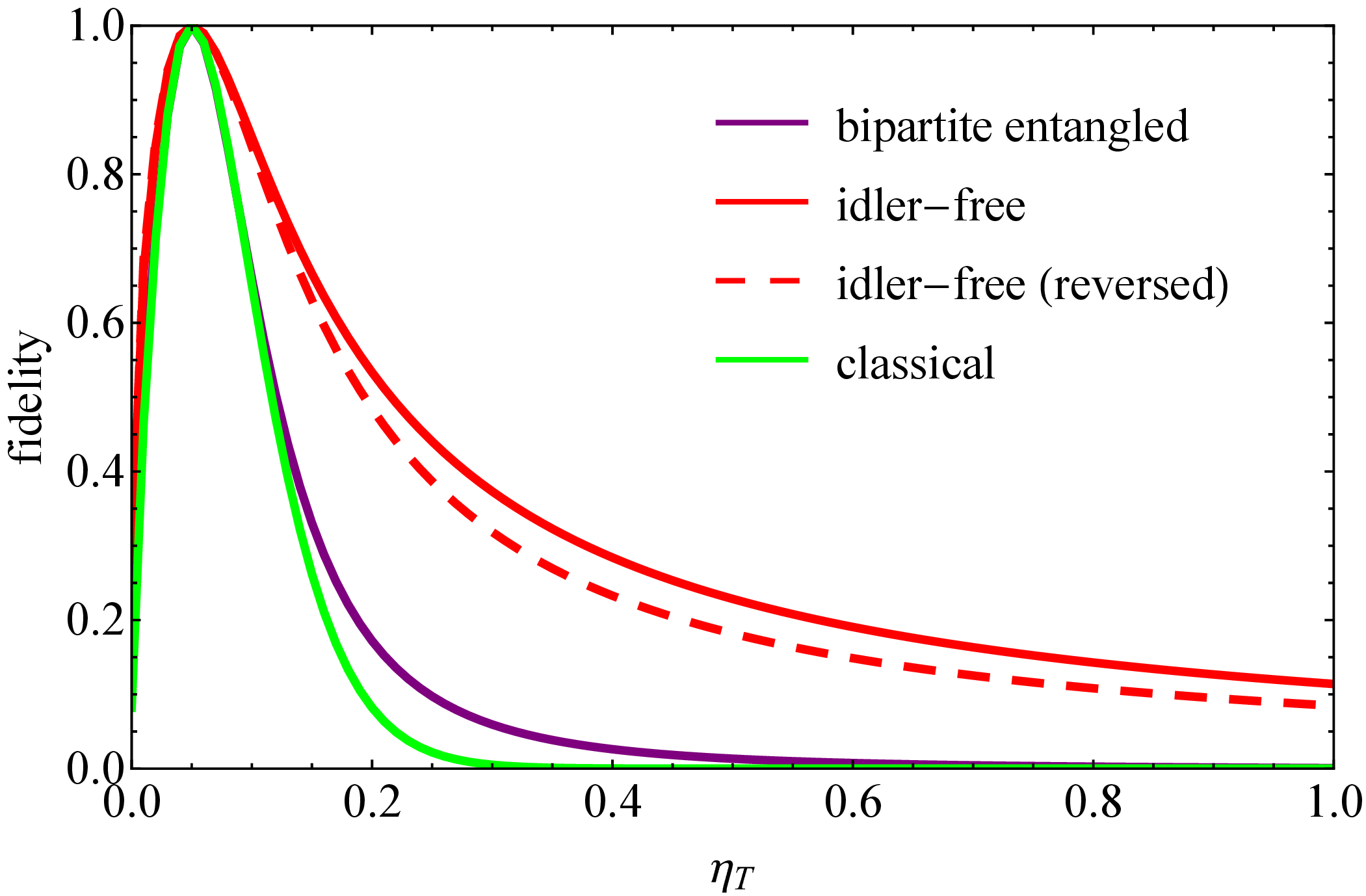}\caption{The output fidelity of the classical, bipartite entangled, and idler-free protocols as a function of the transmissivity of the target channel, $\eta_T$.  We set the transmissivity of the background channels, $\eta_B=0.05$, and impose an energy constraint so that the average number of photons per channel use is no more than $50$. We also set $m=3$, so that there are two identical background channels and one target channel. The output fidelity for the idler-free protocol with $\eta_B$ and $\eta_T$ swapped is also shown. The plot of the output fidelity of the bipartite entangled protocol is the mirror image of the same plot in Fig.~\ref{fig:idler free vs eta}, but the output fidelity of the classical fidelity does not exhibit the same symmetry. The classical protocol has a lower output fidelity than the idler-free protocol across the entire range of $\eta_T$ values and has a lower output fidelity than the bipartite entangled protocol over almost the entire parameter range (with the only exception being a small range of $\eta_T$ values close to $\eta_B$).}
\label{fig:idler free vs eta, low eta}
\end{figure}

We can make this notion more concrete by evaluating the output fidelity functions for the various protocols at the extreme points: $\eta_B=0$ and $\eta_B=1$. We start by observing that both $F^{\mathrm{idler-free},\mathrm{binary}}$ and $F^{\mathrm{bipartite}}$ are unchanged if we simultaneously carry out the replacements $\eta_B\to 1-\eta_B$ and $\eta_T\to 1-\eta_T$. Consequently, these protocols have the same output fidelities for the parameter values $(\eta_B=0,~\eta_T=\epsilon)$ as they do for the parameter values $(\eta_B=1,~\eta_T=1-\epsilon)$. We then calculate
\begin{align}
&F^{\mathrm{idler-free},\mathrm{binary}}_{\eta_B=0,\eta_T=\epsilon}=
F^{\mathrm{idler-free},\mathrm{binary}}_{\eta_B=1,\eta_T=1-\epsilon}=\frac{1}{1+N_S\epsilon},\\
&F^{\mathrm{bipartite}}_{\eta_B=0,\eta_T=\epsilon}=
F^{\mathrm{bipartite}}_{\eta_B=1,\eta_T=1-\epsilon}=\frac{1}{(1+N_S(1-\sqrt{1-\epsilon}))^2}.
\end{align}
These can be expanded (in terms of $\epsilon$) to give
\begin{align}
&F^{\mathrm{idler-free},\mathrm{binary}}_{\eta_B=0(1),\eta_T=\epsilon(1-\epsilon)}\approx 1-N_S\epsilon+N_S^2\epsilon^2+\mathcal{O}(\epsilon^3),\\
&\begin{aligned}
F^{\mathrm{bipartite}}_{\eta_B=0(1),\eta_T=\epsilon(1-\epsilon)}\approx &1-N_S\epsilon+\frac{N_S(3N_S-1)}{4}\epsilon^2\\&+\mathcal{O}(\epsilon^3).
\end{aligned}
\end{align}
Then, for the classical case, which does not have the same symmetry with respect to the transmissivities, we write
\begin{align}
&F^{\mathrm{class}}_{\eta_B=0,\eta_T=\epsilon}=e^{-N_S \epsilon},\\
&F^{\mathrm{class}}_{\eta_B=1,\eta_T=1-\epsilon}=e^{-N_S (1-\sqrt{1-\epsilon})^2}.
\end{align}
Our expansions are
\begin{align}
&F^{\mathrm{class}}_{\eta_B=0,\eta_T=\epsilon}\approx 1-N_S\epsilon+\frac{N_S^2}{2}\epsilon^2+\mathcal{O}(\epsilon^3),\\
&F^{\mathrm{class}}_{\eta_B=1,\eta_T=1-\epsilon}\approx 1-\frac{N_S}{4}\epsilon^2+\mathcal{O}(\epsilon^3).
\end{align}

From these expansions, we can immediately see that in the high transmissivity ($\eta_B=1$) and small $\epsilon$ case, the output fidelity for the classical protocol is higher than for the (binary) idler-free and bipartite entangled protocols, since it does not have a linear term in $\epsilon$. In fact, for the same reason (and using the condition in Eq.~(\ref{adv cond})), there will always be some sufficiently small value of $\epsilon$ for which we will be able to prove a quantum advantage for some sufficiently large number of probes, $M$. In the low transmissivity case ($\eta_B=0$), we have the opposite situation. We can calculate
\begin{align}
&F^{\mathrm{idler-free},\mathrm{binary}}_{\eta_B=0,\eta_T=\epsilon}-F^{\mathrm{class}}_{\eta_B=0,\eta_T=\epsilon}\approx \frac{N_S^2}{2}\epsilon^2+\mathcal{O}(\epsilon^3),\\
&F^{\mathrm{bipartite}}_{\eta_B=0,\eta_T=\epsilon}-F^{\mathrm{class}}_{\eta_B=0,\eta_T=\epsilon}\approx \frac{N_S(N_S-1)}{4}\epsilon^2+\mathcal{O}(\epsilon^3).
\end{align}
Thus, for $\eta_B=0$ and small $\epsilon$, the output fidelity of the classical protocol is always lower than the output fidelity of the idler-free protocol (as expected) and is lower than the output fidelity of the bipartite entangled protocol if the mean number of photons sent into each box per channel use is greater than $1$.

This asymmetric behaviour of the output fidelity for the classical protocol is demonstrated in Fig.~\ref{fig:idler free vs eta, low eta}, where we use the same parameter values that were used in Fig.~\ref{fig:idler free vs eta}, but with the replacement $\eta_B\to 1-\eta_B$ (i.e. $\eta_B=0.05$). In this case, we see that the classical fidelity outperforms (in terms of output fidelities) both the idler-free and the bipartite entangled protocols over almost the entire range of $\eta_T$ values.

\begin{figure}[tb]
\centering
\includegraphics[width=1\linewidth]{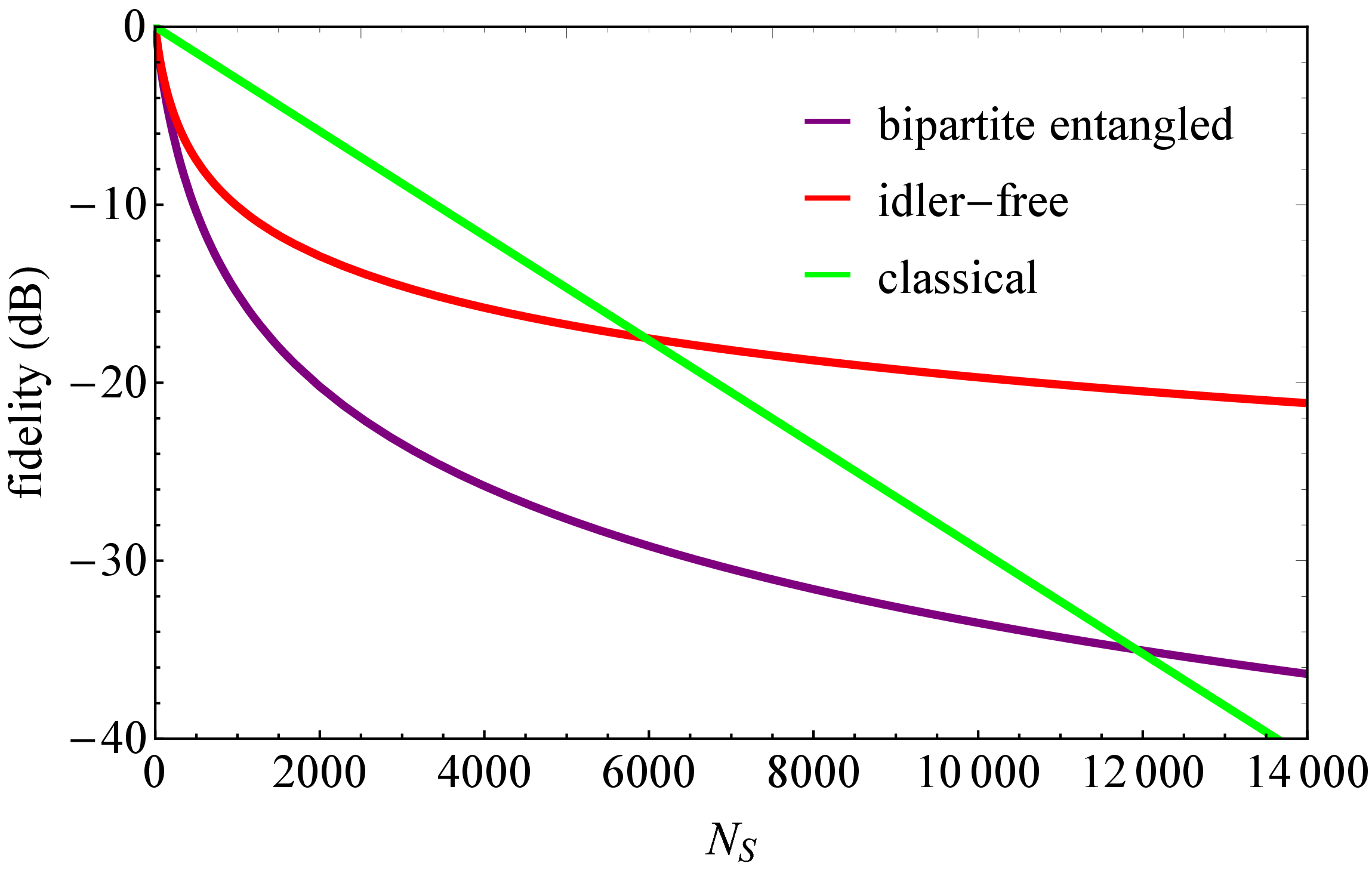}\caption{The output fidelity of the classical, bipartite entangled, and idler-free protocols as a function of the average number of photons in the signal states, $N_S$. We set the background transmissivity, $\eta_B=0.9$, the transmissivity of the target channel, $\eta_T=0.95$, and the number of channels in the sequence, $m=2$. Fidelity is given in decibels. The output fidelity of the classical protocol gives a straight line because the scale is logarithmic and the classical output fidelity scales exponentially. This line crosses the curves representing the output fidelities for both the idler-free and the bipartite entangled protocols, showing that the classical protocol gives a lower output fidelity than either of the other protocols over some parameter ranges.}
\label{fig:idler free vs NS}
\end{figure}

\subsection{Behaviour with respect to the mean number of photons probing each channel}

Fig.~\ref{fig:idler free vs NS} plots the output fidelities against the average number of photons sent into each channel. We have set $\eta_B=0.9$, $\eta_T=0.95$, and $m=2$; since there are only two channels in the sequence, switching $\eta_B$ and $\eta_T$ does not result in a different task, and so we do not plot the case with $\eta_B$ and $\eta_T$ switched. The fidelity is given in decibels; this allows it to be clearly seen that the output fidelity of the classical protocol scales exponentially with $N_S$, since the curve is linear in a log scale. In fact, this is evident from the form of the expression in Eq.~(\ref{eq: fid_class}). On the other hand, $F^{\mathrm{bipartite}}$ is polynomial in $N_S$. Considering the expression in Eq.~(\ref{eq: bipartite ent fid}) for large $N_S$, we see that it scales as roughly $N_S^{-2}$. We can see from Fig.~\ref{fig:idler free vs NS} that the scaling of the idler-free output fidelity is also less than exponential. From Eq.~(\ref{eq: idler free binary fid}), it can be seen that the output fidelity in the $m=2$ case scales as approximately $N_S^{-1}$ for large $N_S$. Since the output fidelity is lowest in the $m=2$ case, the classical protocol will always beat the idler-free protocol (and the bipartite entangled protocol) for sufficiently high $N_S$, due to the different scalings.

\section{Mixed strategy}

The classical protocol uses coherent state probes, which have non-zero first moments and have the identity matrix for their CMs. The idler-free protocol uses probe systems that have no non-zero first moments but that have non-trivial CMs. We will now define what we call the mixed strategy as a natural combination of these two strategies. This protocol uses probe states that have both non-zero first moments and non-trivial CMs. More specifically, we produce our probe state, $\rho_{\mathrm{mixed}}$, by displacing the state used in the idler-free protocol, $\rho_{\mathrm{in}}$ (with its CM as defined in Eq.~(\ref{eq: idler-free input})), using the displacement operator
\begin{equation}
D(\boldsymbol{\alpha})=d_0(\alpha)\otimes d_1(\alpha)\otimes\ldots d_{m-1}(\alpha).
\end{equation}
In other words, every mode is subject to the same displacement, $d(\alpha)$.

\begin{figure}[tb]
\centering
\includegraphics[width=1\linewidth]{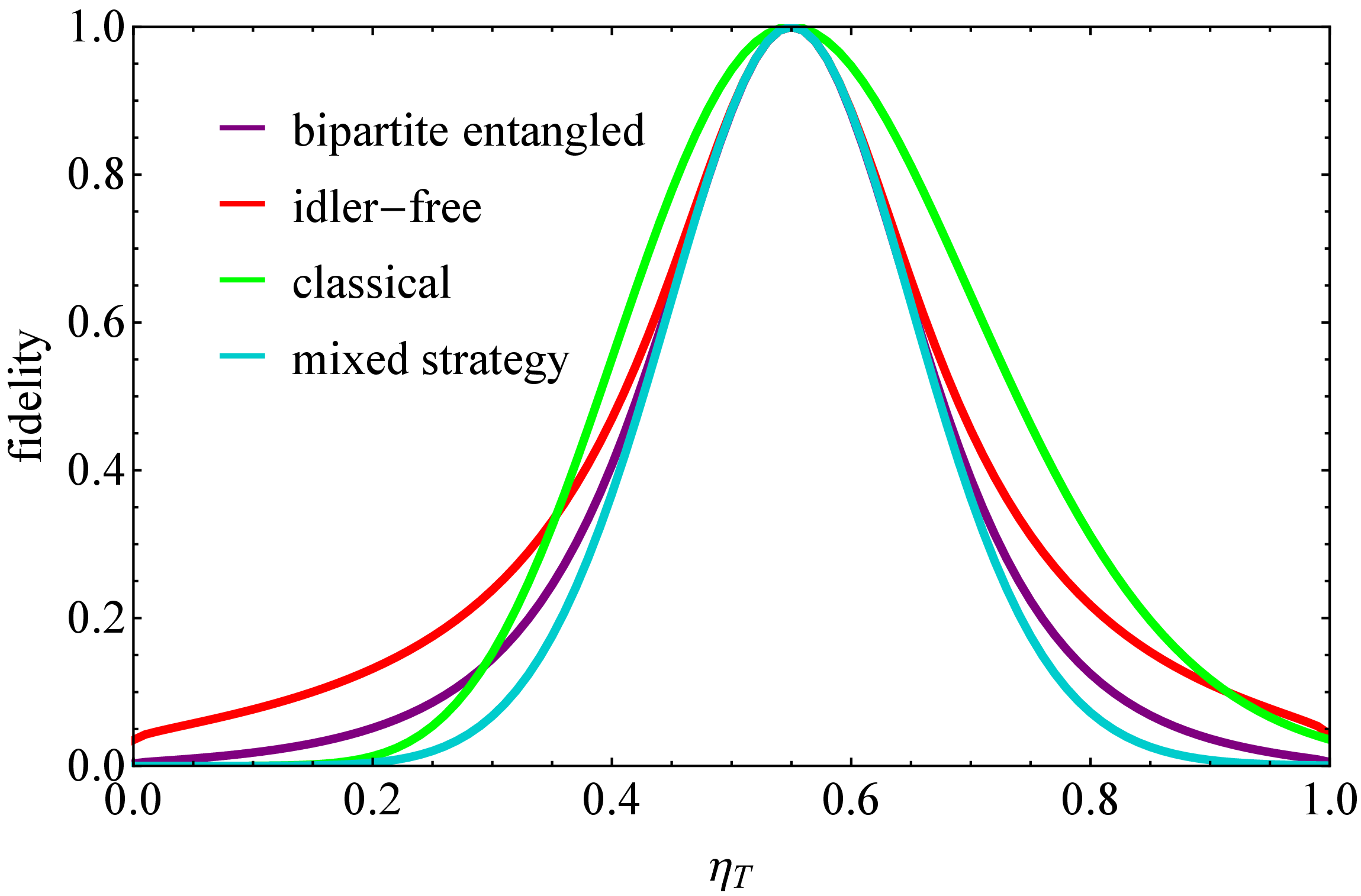}\caption{The output fidelity of the classical, bipartite entangled, idler-free, and mixed strategy protocols as a function of the transmissivity of the target channel, $\eta_T$.  We set the transmissivity of the background channel, $\eta_B=0.55$, and impose an energy constraint so that the average number of photons per channel use is no more than $50$. We set $m=2$, so that there is only one background channel and one target channel. The mixed strategy improves on all of the other strategies (in terms of output fidelity) over almost the entire range of $\eta_T$ values. It is able to beat the bipartite entangled protocol even in a parameter range in which neither of the two extremal strategies (the classical protocol and the idler-free protocol) beat it.}
\label{fig:mixed strategy}
\end{figure}

Since the displacement increases the mean photon numbers of the signal modes, we must divide the available energy between displacing the states and correlating the modes (i.e. squeezing) in order to meet the energy constraint. Each box sees a thermal coherent state~\cite{oz-vogt_thermal_1991} and so we must calculate the mean number of photons accordingly. For such a state, the mean number of photons is the sum of the mean number of photons for the corresponding thermal state (the state prior to displacement) and the mean number of photons for the corresponding coherent state (the state obtained if the same displacement were applied to a vacuum state). Accordingly, the energy can simply be divided by some mixing parameter $\kappa$, so that $\kappa$ is the proportion of the available energy that is used to entangle the signal modes and $(1-\kappa)$ is the proportion of the energy that is used to displace the states. Alternatively, $\kappa N_S$ is the mean number of photons per mode before the probe is displaced.

The CM of our new probe state is the CM given in Eq.~(\ref{eq: idler-free input}) but with $\mu$ replaced by $\mu'$, where
\begin{equation}
\mu'=2\kappa N_S+1=\kappa(\mu-1)+1.
\end{equation}
The first moments of the probe state (up to some arbitrary phase rotation on each individual mode) are
\begin{align}
\boldsymbol{x}&=(q_0,p_0,q_1,p_1,\ldots q_{m-1},p_{m-1})^T\\
&=2\sqrt{(1-\kappa)N_S}(1,0,1,0,\ldots 1,0)^T.
\end{align}
We can then choose the value of $\kappa$ that minimises the output fidelity (e.g. by numerical minimisation). Note that if $\kappa=0$, we have the classical protocol and if $\kappa=1$, we have the idler-free protocol, hence we can always perform at least as well as the best of the two strategies.

Fig.~\ref{fig:mixed strategy} plots the output fidelities of the various protocols, including the mixed strategy, against the transmissivity of the target channel, $\eta_T$.  We fix the background transmissivity, $\eta_B=0.55$, the number of channels in the sequence, $m=2$, and the average number of photons sent through each channel per channel use, $N_S=50$. In terms of the output fidelity, the mixed strategy outperforms all of the other protocols over almost the entire parameter range. It is not surprising that it is never worse than either the classical protocol or the idler-free protocol: this is by construction, since we can always choose $\kappa=0(1)$ if the classical (idler-free) protocol is better than any mixed strategy with $\kappa>0$ ($\kappa<1$). More interesting is the fact that the mixed strategy often achieves a significantly better output fidelity than either of the extremal protocols and is able to beat the bipartite entangled protocol even for values of $\eta_T$ for which neither the classical protocol nor the idler-free protocol do so. In fact, for $\eta_T\gtrapprox0.86$, we are even able to prove an advantage in terms of discrimination error probability (using the condition in Eq.~(\ref{adv cond})) for the mixed strategy over both the classical and the idler-free strategies. Recall also that the mixed strategy does not require an idler and so still has the benefit of not requiring a quantum memory.

\section{Application to position-based quantum reading}

\begin{figure}[tb]
\centering
\includegraphics[width=1\linewidth]{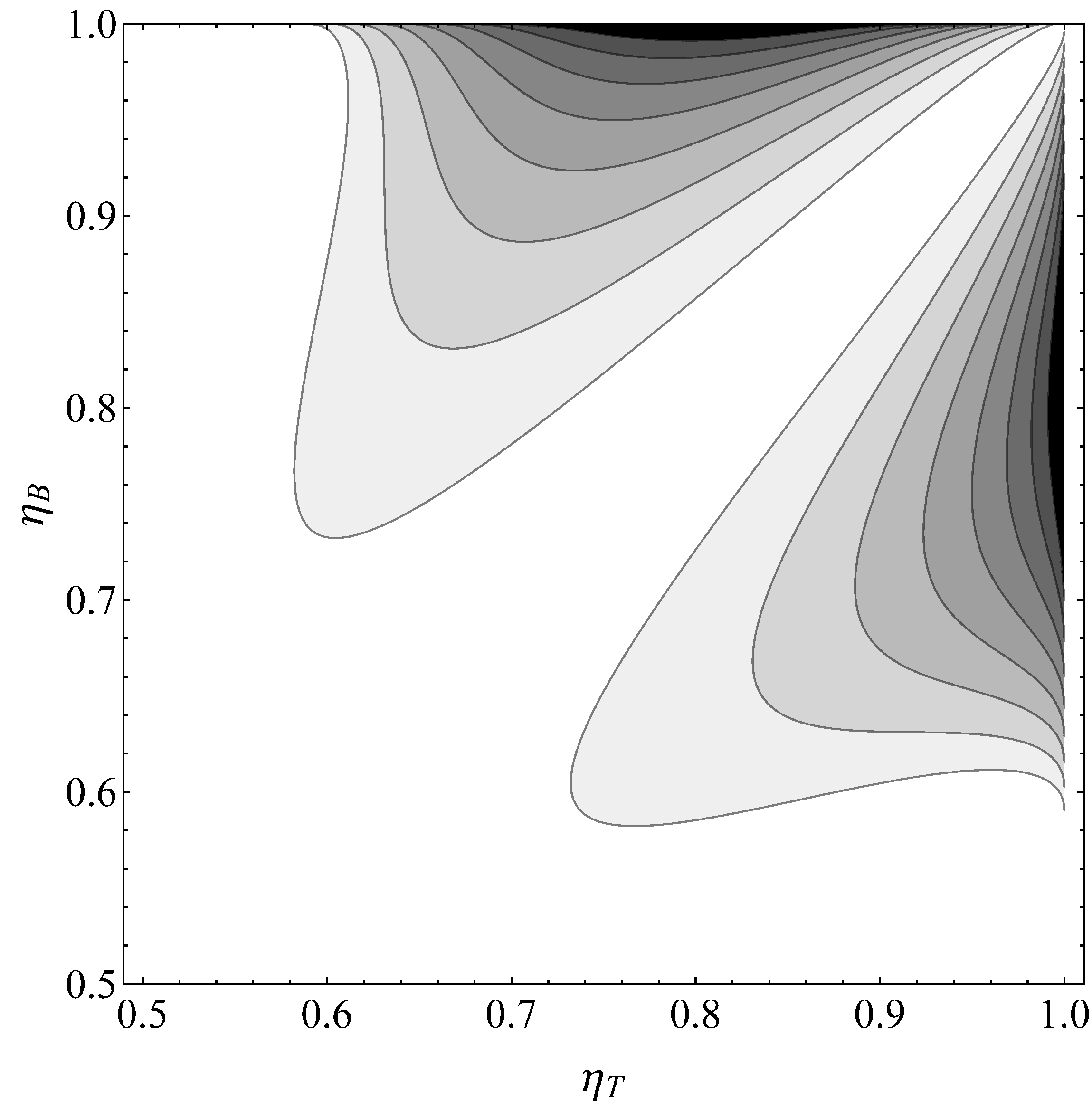}\caption{The region in which we can demonstrate quantum advantage for quantum reading of a bit encoded in $m=2$ memory cells when probing $M=20$ times with $N_S=20$ photons per channel per probe. The advantage is expressed as the log (base 10) of the ratio between the upper bound on the error probability for the idler-free protocol and the lower bound on the error probability for the classical protocol, with each contour line representing a change of $1$ and the outermost line representing $0$. We are able to prove quantum advantage for high transmissivities of both the target and the background. Note the symmetry about the line $\eta_B=\eta_T$.}
\label{fig:reading idler-free}
\end{figure}

We now combine the various fidelities with the error bounds in Eqs.~(\ref{eq: error prob UB}) and (\ref{eq: error prob LB}) in order to demonstrate the regions in which we are able to show a quantum advantage for the idler-free protocol and the mixed strategy. As an example, we consider the position-based quantum reading model used in Ref.~\cite{zhuang_entanglement-enhanced_2020}, in which memory cells have one of two possible reflectivities (determined by whether the cell is a target or a background cell), and classical information is stored in the position of a target cell amongst background cells.

Zhuang and Pirandola generally considered high values of $m$ ($~100$), however, for the idler-free protocol, this is not practicable because output fidelity increases with $m$, removing any quantum advantage for large $m$. Instead, we consider scenarios with $2$ or $3$ cells.

We start by considering a scenario with $m=2$ cells, probed using $N_S=20$ photons per channel use and with a total of $M=20$ uses of the channel sequence (see Fig.~\ref{fig:reading idler-free}). Here we can see that we get quantum advantage when the transmissivities of both the target and the background channels are high but are not too similar. For the ideal case, in which the transmissivity of the background channel is $1$, we can prove a quantum advantage for any $\eta_T\gtrapprox 0.59$.

\begin{figure}[tb]
\centering
\includegraphics[width=1\linewidth]{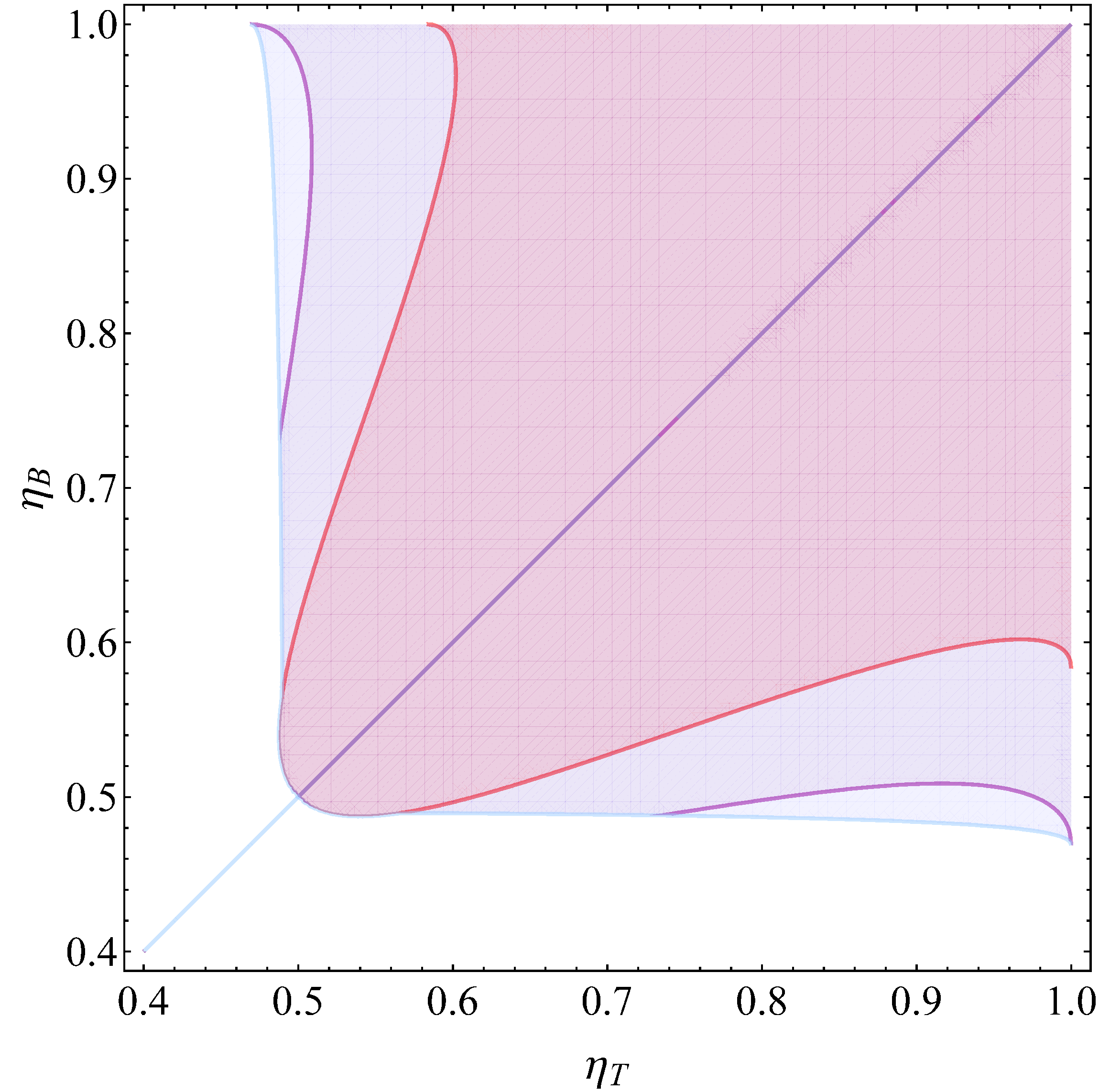}\caption{The region in which we can demonstrate quantum advantage for quantum reading of a bit encoded in $m=2$ memory cells, for some number of probes, using various strategies, with $N_S=20$ photons per channel per probe. The protocols considered are the idler-free protocol (innermost shaded region), the bipartite entangled protocol (middle shaded region), and the mixed strategy (outermost shaded region). We plot the regions in which each strategy is able to achieve a quantum advantage over the best classical strategy.}
\label{fig:reading strategies}
\end{figure}

Next, we consider a scenario with $m=2$ cells, again probed using $N_S=20$ photons per channel use, and compare the performance of the various strategies. Specifically, in Fig.~\ref{fig:reading strategies}, we plot the regions in which we can show a quantum advantage, for some number of probes, for the idler-free, bipartite entangled, and mixed strategies. These regions are defined using Eq.~(\ref{adv cond}); we therefore do not set $M$ as we are simply determining whether there exists $M$ such that we can prove a quantum advantage. We see that, in this regime, the mixed strategy is able to show a quantum advantage over a range of parameter values for which the bipartite entangled protocol cannot. The idler-free strategy is the least powerful, in this sense, but is still able to show a quantum advantage when the transmissivities of the target and background channels are high.

\begin{figure}[tb]
\centering
\includegraphics[width=1\linewidth]{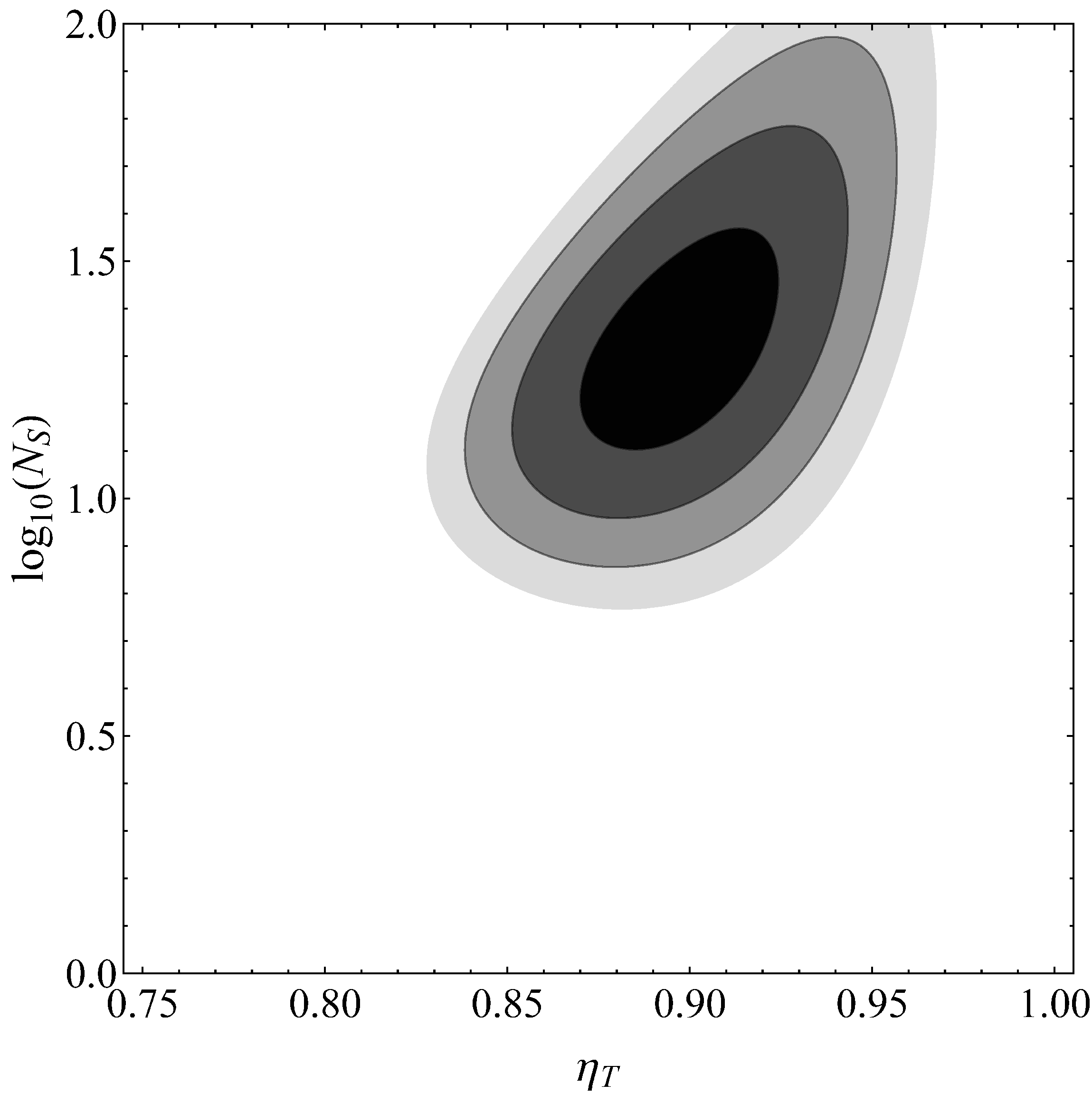}\caption{The region in which we can demonstrate quantum advantage for quantum reading of a bit encoded in $m=3$ memory cells when probing with $m M N_S=1800$ photons in total. The advantage is expressed as the log (base 10) of the ratio between the upper bound on the error probability for the idler-free protocol and the lower bound on the error probability for the classical protocol, with each contour line representing a change of $0.2$ and the outermost line representing $0$. We are able to prove quantum advantage for target transmissivities between $~0.83$ and $~0.97$ and numbers of photons per channel use greater than about $6$.}
\label{fig:reading fixed MN}
\end{figure}

Note that we are always comparing the upper bound on the error probability of the quantum strategy to the lower bound on the error probability of the classical strategy. In fact, there may be some specific receiver design (measurement) that allows a quantum protocol to achieve a lower error rate than is given by the upper bound in Eq.~(\ref{eq: error prob UB}). For the bipartite entangled protocol, such a receiver design is known: the CN receiver of Ref.~\cite{zhuang_entanglement-enhanced_2020,shi_entanglement-assisted_2020}. Over some parameter ranges, the error probability given by the CN receiver is smaller than the one given by Eq.~(\ref{eq: error prob UB}) (based on the PGM).

Finally, we consider the ideal scenario in which the background channel has a transmissivity of $\eta_B=1$. In this case, we constrain the total number of photons used for probing by fixing $MN_S=600$. Then, as is done in Ref.~\cite{zhuang_entanglement-enhanced_2020}, we plot the quantum advantage as a function of $\eta_T$ and $N_S$ (see Fig.~\ref{fig:reading fixed MN}). We set $m=3$ and observe that the range of values over which we can prove a quantum advantage is much smaller than for $m=2$. In fact, we only see a quantum advantage for $\eta_T$ close to $1$ and for a large number of photons per channel use.

It is to be noted that not being able to prove a quantum advantage is not a proof that there is no quantum advantage. Unless we can prove an advantage the other way (using the condition in Eq.~(\ref{adv cond}), for example), we may still have an advantage with some strategy paired with a particular choice of receiver. As mentioned previously, the CN receiver gives an error probability for the bipartite entangled case that is much tighter than the upper boundon the error from Eq.~(\ref{eq: error prob UB}), and there could be some measurement for the idler-free protocol or the mixed strategy that offers a similar improvement in the upper bound on the error probability. This could be a avenue for future research.

\section{Conclusion}

We have presented a type of protocol for channel position finding that does not involve idlers and so does not require a quantum memory in order to work. The source used remains non-classical and shows an advantage over classical protocols for channel position finding on a sequence of pure loss channels over a range of parameter values and especially for high transmissivities. We have investigated the behaviour of the output fidelities of the various types of protocols over different transmissivities, numbers of channels, and numbers of photons. We then developed an intermediate between the idler-free and classical protocols, which we called the mixed strategy. This new protocol is, by design, always at least as good (in the sense of having an equally low error probability) as the best of the two constituent protocols and, in fact, is often significantly better than either. The protocols were then applied to position-based quantum reading, and their performances were investigated.

Future research could focus on finding specific receivers for the idler-free protocol and the mixed strategy, in order to improve their error probabilities. The probe used for the idler-free protocol (and the mixed strategy) is not pure (for $m>2$). Another open question is whether there exists some pure state that could offer a better performance than a probe with its CM as defined in Eq.~(\ref{eq: idler-free input}). Giving the idler-free probe non-zero first moments greatly boosted its performance. Since the bipartite entangled protocol outperformed the idler-free protocol, it is possible that a displaced TMSV state could perform even better than the mixed strategy that we have presented (whilst no longer being idler-free).

\smallskip
\textbf{Acknowledgments.}~This work was funded by the European Union via ``Quantum readout techniques and technologies'' (QUARTET, Grant agreement No 862644) and via Continuous Variable Quantum Communications” (CiViQ, Grant agreement No 820466), and by the EPSRC via the Quantum Communications Hub (Grants No. EP/M013472/1 and No. EP/T001011/1). L.B. acknowledges support by the program ``Rita Levi Montalcini'' for young researchers. Q.Z. acknowledges the Defense Advanced Research Projects Agency (DARPA) under Young Faculty Award (YFA) Grant No. N660012014029 and Craig M. Berge Dean's Faculty Fellowship of University of Arizona.

\end{document}